\documentclass[12pt,journal,onecolumn]{IEEEtran}

\usepackage{amssymb,amsmath,amsfonts,bm,epsfig,graphicx,theorem,latexsym}
\usepackage{rotating,setspace,latexsym,epsf,color,subfigure,epstopdf}
\usepackage{cite,authblk,algorithmic}

\theorembodyfont{\normalfont}
{\theoremstyle{plain}
\newtheorem{theorem}{Theorem}
\newtheorem{lemma}{Lemma}

\newenvironment{Proof}[1]{\medskip\par\noindent
{\bf Proof:\,}\,#1}{{\mbox{\,$\blacksquare$}\par}}

\setstretch{1.6}
\textwidth 6.5 in
       \oddsidemargin 0 in
       \evensidemargin 0 in
       \textheight 9.1 in
       \topmargin -0.5 in
       
\begin{document}
\IEEEoverridecommandlockouts

\title{Transmission with Energy Harvesting Nodes in Fading Wireless Channels: Optimal Policies\thanks{Omur Ozel and Sennur Ulukus are with the Department of Electrical and Computer Engineering, University of Maryland, College Park, MD 20742, USA. Kaya Tutuncuoglu and Aylin Yener are with the Department of Electrical Engineering, Pennsylvania State University, University Park, PA 16802, USA. Jing Yang was with the Department of Electrical and Computer Engineering, University of Maryland, College Park, MD 20742, USA. She is now with the Department of Electrical and Computer Engineering, University of Wisconsin-Madison, WI 53706, USA. This work was supported by NSF Grants CNS 09-64632, CNS 09-64364 and presented in part at the Conference on Information Sciences and Systems (CISS), Baltimore, MD, March 2011 and at IEEE International Conference on Computer Communications (INFOCOM), Shangai, China, April 2011.}}

\author{Omur Ozel, Kaya Tutuncuoglu, Jing Yang, Sennur Ulukus and Aylin Yener}
\maketitle\thispagestyle{empty}

\vspace{-0.4in}

\begin{abstract}
Wireless systems comprised of rechargeable nodes have a significantly prolonged lifetime and are sustainable. A distinct characteristic of these systems is the fact that the nodes can harvest energy throughout the duration in which communication takes place. As such, transmission policies of the nodes need to adapt to these harvested energy arrivals. In this paper, we consider optimization of point-to-point data transmission with an energy harvesting transmitter which has a limited battery capacity, communicating in a wireless fading channel. We consider two objectives: maximizing the throughput by a deadline, and minimizing the transmission completion time of the communication session. We optimize these objectives by controlling the time sequence of transmit powers subject to energy storage capacity and causality constraints. We, first, study optimal offline policies. We introduce a {\it directional water-filling} algorithm which provides a simple and concise interpretation of the necessary optimality conditions. We show the optimality of an adaptive directional water-filling algorithm for the throughput maximization problem. We solve the transmission completion time minimization problem by utilizing its equivalence to its throughput maximization counterpart. Next, we consider online policies. We use stochastic dynamic programming to solve for the optimal online policy that maximizes the average number of bits delivered by a deadline under stochastic fading and energy arrival processes with causal channel state feedback. We also propose near-optimal policies with reduced complexity, and numerically study their performances along with the performances of the offline and online optimal policies under various different configurations.
\end{abstract}
\begin{keywords} Energy harvesting, rechargeable wireless networks, throughput maximization, transmission completion time minimization, directional water-filling, dynamic programming. \end{keywords}

\newpage
\pagestyle{plain}
\setcounter{page}{1}
\pagenumbering{arabic}

\section{Introduction}
This paper considers wireless communication using energy harvesting
transmitters. In such a scenario, incremental energy is harvested by
the transmitter during the course of data transmission from random
energy sources. As such, energy becomes available for
packet transmission at random times and in random amounts. In
addition, the wireless communication channel fluctuates randomly due
to fading. These together lead to a need for designing new
transmission strategies that can best take advantage of and adapt to
the random energy arrivals as well as channel variations in time.

The simplest system model for which this setting leads to new
design insights is a wireless link with a rechargeable
transmitter, which we consider here. The incoming energy can be
stored in the battery of the rechargeable transmitter for
future use. However, this battery has finite storage capacity
and the transmission policy needs to guarantee that there is
sufficient battery space for each energy arrival, otherwise
incoming energy cannot be saved and will be wasted. In this setting, we find optimal offline and online transmission schemes that adapt the instantaneous transmit power to the variations in the energy and fade levels.

There has been recent research effort on understanding data transmission with an energy harvesting transmitter that has a rechargeable battery \cite{yates09TWC,sharma10TWC, tassiulas10TWC,Zhang10ISIT,jing10ciss,tcom-submit,kaya_subm}. In~\cite{yates09TWC}, data transmission with energy harvesting sensors is considered, and the optimal online policy for controlling admissions into the data buffer is derived using a dynamic programming framework. In~\cite{sharma10TWC}, energy management policies which stabilize the data queue are proposed for single-user communication and under a linear approximation, some delay optimality properties are derived. In~\cite{tassiulas10TWC}, the optimality of a variant of the back pressure algorithm using energy queues is shown. In~\cite{Zhang10ISIT}, throughput optimal energy allocation is studied for energy harvesting systems in a time constrained slotted setting. In~\cite{jing10ciss,tcom-submit}, minimization of the transmission completion time is considered in an energy harvesting system and the optimal solution is obtained using a geometric framework similar to the calculus approach presented in \cite{eytan09}. In~\cite{kaya_subm}, energy harvesting transmitters with batteries of finite energy storage capacity are considered and the problem of throughput maximization by a deadline is solved in a static channel. 

An earlier line of research considered the problem of energy management in communications satellites~\cite{fu03TON,alvinfu06TWC}. In~\cite{fu03TON}, various energy allocation problems in solar powered communication satellites are solved using dynamic programming. In~\cite{alvinfu06TWC}, optimal energy allocation to a fixed number of time slots is derived under time-varying channel gains and with offline and online knowledge of the channel state at the transmitter. Another related line of research considered energy minimal transmission problems with deadline constraints~\cite{elif01INFOCOM, it_2004, eytan09, zafer05alerton}. Our work provides optimal transmission policies to maximize the throughput and minimize the transmission completion time, under channel fluctuations and energy variations, in a continuous time model, generalizing these related works, \cite{Zhang10ISIT,alvinfu06TWC,jing10ciss, tcom-submit, kaya_subm,elif01INFOCOM, it_2004, eytan09, zafer05alerton}, from various different perspectives.

In particular, we consider two related optimization problems. The first problem is the maximization of the number of bits (or throughput) transmitted by a deadline $T$. The second problem is the minimization of the time (or delay) by which the transmission of $B$ bits is completed. We solve the first problem under deterministic (offline) and stochastic (online) settings, and we solve the second problem in the deterministic setting. We start the analysis by considering the first problem in a static channel under offline knowledge. The solution calls for a new algorithm, termed {\it directional water-filling}. Taking into account the causality constraints on the energy usage, i.e., the energy can be saved and used in the future, the algorithm allows energy flow only to the right. In the algorithmic implementation of the solution, we utilize {\it right permeable taps} at each energy arrival point. This solution serves as a building block for the fading case. Specifically, we show that a directional water-filling algorithm that adapts to both energy arrivals and channel fade levels is optimal. Next, we consider the second problem, i.e., the minimization of the time by which transmission of $B$ bits is completed. We use the solution of the first problem to solve this second problem. This is accomplished by mapping the first problem to the second problem by means of the {\it maximum departure curve}. This completes the identification of the optimal offline policies in the fading channel.

Next, we set out to find online policies. We address online scheduling for maximum throughput by the deadline $T$ in a setting where fading level changes and energy arrives as random processes in time. Assuming statistical knowledge and causal information of the energy and fading variations, we solve for the optimal online power policy by using continuous time stochastic dynamic programming \cite{zafer05alerton,BersekasDynProg}. To reduce the complexity required by the dynamic programming solution, we propose simple online algorithms that perform near-optimal. Finally, we provide a thorough numerical study of the proposed algorithms under various system settings. 

\section{System Model} 
\label{model}
\begin{figure}[t]
\begin{center}
\includegraphics[width=0.6\linewidth]{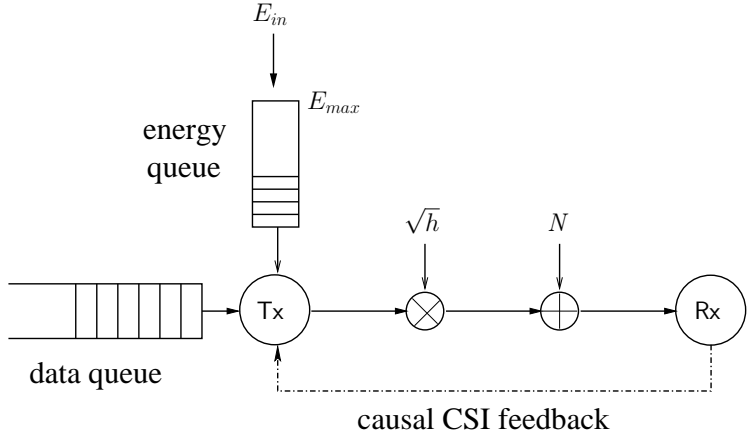}
\end{center}
\caption{Additive Gaussian fading channel with an energy harvesting transmitter and causal channel state information (CSI) feedback.}
\label{mod1}
\end{figure}

We consider a single-user fading channel with additive Gaussian noise and causal channel state information (CSI) feedback as shown in Fig.~\ref{mod1}. The transmitter has two queues, the data queue where
data packets are stored, and an energy queue where the arriving
(harvested) energy is stored. The energy queue, i.e., the battery, can
store at most $E_{max}$ units of energy, which is used
only for transmission, i.e., energy required for processing is not considered.

The received signal $y$ is given by
$y=\sqrt{h}x + n$, where $h$ is the (squared) fading, $x$ is the
channel input, and $n$ is a Gaussian random noise with zero-mean
and unit-variance. Whenever an input signal $x$ is transmitted with power $p$ in an epoch
of duration $L$, $\frac{L}{2} \log\left(1 + hp\right)$ bits of data is
served out from the backlog with the cost of $Lp$ units of energy
depletion from the energy queue. This follows from the Gaussian channel capacity formula. The bandwidth is sufficiently wide so that $L$ can take small values and we approximate the slotted system to a continuous time system. Hence, we say that if at time $t$ the transmit power of the signal is $x^2(t)=p(t)$, the instantaneous rate $r(t)$ in bits per channel use is
\begin{align}
\label{rate}
r(t)=\frac{1}{2}\log\left(1 + h(t)p(t)\right)
\end{align}

\begin{figure}[t]
\begin{center}
\includegraphics[width=0.7\linewidth]{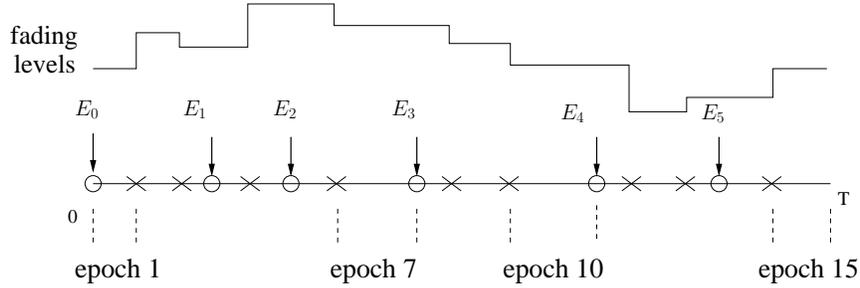}
\end{center}
\caption{The system model and epochs under channel fading.}
\label{mod}
\end{figure}

Following a model similar to \cite{it_2004}, we assume that the fading level $h$ and energy arrivals are stochastic processes in time that are marked by Poisson counting processes with rates $\lambda_h$ and $\lambda_e$, respectively. Therefore, changes in fading level and energy arrivals occur in countable time instants, which are indexed respectively as $t_1^{f},t_2^{f},\ldots,t_{n}^{f},\ldots$ and $t_1^{e},t_2^{e},\ldots,t_{n}^{e},\ldots$ with the convention that $t_1^{e}=t_1^{f}=0$. By the Poisson property, the inter-occurrence times $t_i^f-t_{i-1}^f$ and $t_j^e-t_{j-1}^e$ are exponentially distributed with means $1/\lambda_f$ and $1/\lambda_e$, respectively. The fading level in $[0,t_1^{f})$ is $h_1$, in $[t_1^{f},t_2^{f})$ is $h_2$, and so on. Similarly, $E_i$ units of energy arrives at time $t_i^{e}$, and $E_0$ units of energy is available at time $0$. Hence $\{(t_i^e,E_i)\}_{i=0}^\infty$ and $\{(t_i^f,h_i)\}_{i=1}^\infty$ completely define the events that take place during the course of data transmission. This model is shown in Fig.~\ref{mod}. The incoming energy is first buffered in the battery before it is used in data transmission, and the transmitter is allowed to use the battery energy only. Accordingly, we assume $E_i \leq E_{max}$ for all $i$ as otherwise excess energy cannot be accommodated in the battery anyway. 

In the sequel, we will refer to a change in the channel fading level or in the energy level as an {\it event} and the time interval between two consecutive events as an {\it epoch}. More precisely, epoch $i$ is defined as the time interval $[t_i,t_{i+1})$ where $t_i$ and $t_{i+1}$ are the times at which successive events happen and the length of the epoch is $L_i=t_{i+1}-t_i$. Naturally, energy arrival information is causally available to the transmitter. Moreover, by virtue of the causal feedback link, perfect information of the channel fade level is available to the transmitter. Therefore, at time $t$ all $\{E_i\}$ and $\{h_j\}$ such that $t_i^e<t$ and $t_j^f<t$ are known perfectly by the transmitter.

A power management policy is denoted as $p(t)$ for $t \in [0,T]$. There are two constraints on $p(t)$, due to energy arrivals at random times and also due to finite battery storage capacity. Since energy that has not arrived yet cannot be used at the current time, there is a causality constraint on the power management policy as:
\begin{align} 
\label{main_causal} 
\int_0^{t_i^e} p(u)du \leq \sum_{j=0}^{i-1} E_j, \quad \forall i 
\end{align}
where the limit of the integral $t_i^e$ should be interpreted as ${t_i}^e-\epsilon$, for small enough $\epsilon$. Moreover, due to the finite battery storage capacity, we need to make sure that energy level in the battery never exceeds $E_{max}$. Since energy arrives at certain time points, it is sufficient to ensure that the energy level in the battery never exceeds $E_{max}$ at the times of energy arrivals. Let $d(t)= \max\{ t^e_i: t^e_i \leq t\}$. Then, 
\begin{align} 
\label{main_max} 
\sum_{j=0}^{d(t)} E_j - \int_0^t p(u) du \leq E_{max}, \quad \forall t \in [0,T]   \end{align}

We emphasize that our system model is continuous rather than slotted. In slotted models,
e.g., \cite{fu03TON,tassiulas10TWC,Zhang10ISIT}, the energy input-output
relationship is written for an entire slot. Such models allow energies
larger than $E_{max}$ to enter the battery and be used for
transmission in a given single slot. Our continuous system model
prohibits such occurrences.  

Our ultimate goal is to develop an online algorithm that determines the transmit power as a function of time using the causal knowledge of the system, e.g., the instantaneous energy state and fading CSI. We will start our development by considering the optimal offline policy. 

\section{Maximizing Throughput in a Static Channel}
In this section, we consider maximizing the number of bits delivered by a deadline $T$, in
a non-fading channel with offline knowledge of energy arrivals which occur at times
$\{t_1,t_2,\ldots,t_N\}$ in amounts
$\{E_1,E_2,\ldots,E_N\}$. The epoch lengths are $L_i=t_{i}
- t_{i-1}$ for $i=1,\ldots,N$ with $t_0 = 0$, and
$L_{N+1}=T-t_N$. There are a total of $N+1$ epochs. The optimization is subject to causality constraints on the harvested energy, and the
finite storage constraint on the rechargeable battery. This problem was solved in \cite{kaya_subm} using a geometric framework. Here, we provide the formulation for completeness and provide an alternative algorithmic solution which will serve as the building block for the solution for the fading channel presented in the next section.

First, we note that the transmit power must be kept constant in each
epoch \cite{jing10ciss, tcom-submit, kaya_subm}, due to the concavity of rate in power. Let us denote the power in
epoch $i$ by $p_i$. The causality constraints in (\ref{main_causal})
reduce to the following constraints on $p_i$,
\begin{align} 
\label{const1} 
\sum_{i=1}^{\ell} L_i p_i \leq \sum_{i=0}^{\ell-1} E_i , \quad \ell=1,\ldots, N+1 \end{align}
Moreover, since the energy level in the battery is the highest at
instants when energy arrives, the battery capacity constraints in
(\ref{main_max}) reduce to a countable number of constraints, as
follows
\begin{align} \label{const2} 
\sum_{i=0}^{\ell} E_i - \sum_{i=1}^{\ell} L_ip_i \leq E_{max}, \quad \ell=1,\ldots,N 
\end{align}
Note that since $E_0>0$, there is no incentive to make $p_i=0$ for any
$i$. Hence, $p_i>0$ is necessary for optimality.

 The
optimization problem is:
\begin{eqnarray}
\label{opt1}
\max_{p_i\geq 0} & & \sum_{i=1}^{N+1} \frac{L_i}{2}\log\left(1 + p_i\right)\\
\mbox{s.t.} & & \sum_{i=1}^{\ell} L_i p_i \leq \sum_{i=0}^{\ell-1} E_i, \quad \ell = 1,\ldots, N+1 \label{opt11}\\
 & & \sum_{i=0}^{\ell}  E_i - \sum_{i=1}^{\ell}  L_ip_i \leq E_{max}, \quad \ell= 1,\ldots, N \label{opt12}
\end{eqnarray}
We note that the constraint in (\ref{opt11}) must be satisfied with
equality for $\ell=N+1$, otherwise, we can always increase some
$p_{i}$ without conflicting any other constraints, increasing the
resulting number of bits transmitted.

Note that the objective function in (\ref{opt1}) is concave in the vector of powers
since it is a sum of $\log$ functions, which are concave themselves. In addition, the constraint set is convex as it is composed of linear constraints. Hence, the above optimization problem is a convex optimization problem, and
has a unique maximizer. We define the following Lagrangian function
\cite{BersekasConvexOpt} for any $\lambda_i \geq 0$ and $\mu_i \geq
0$,
\begin{align}
\mathcal{L}=  \sum_{i=1}^{N+1} \frac{L_i}{2}\log\left(1 + p_i\right) 
- \sum_{j=1}^{N+1} \lambda_j \left( \sum_{i=1}^j L_ip_i - \sum_{i=0}^{j-1}E_i  \right) - \sum_{j=1}^N \mu_j\left(\sum_{i=0}^j E_i - \sum_{i=1}^j L_ip_i - E_{max}\right) \label{lag}
\end{align} 
Lagrange multipliers $\{\lambda_i\}$ are associated with constraints in (\ref{opt11}) and $\{\mu_i\}$ are associated with (\ref{opt12}). Additional complimentary slackness conditions are as follows,
\begin{align} 
\label{cslck1} 
\lambda_j\left(\sum_{i=1}^j L_ip_i - \sum_{i=0}^{j-1}E_i \right) &= 0, \quad j=1,\ldots,N \\
\label{cslck2} 
\mu_j\left(\sum_{i=0}^j E_i - \sum_{i=1}^j L_ip_i - E_{max}\right) &= 0, \quad j=1,\ldots,N 
\end{align}
In (\ref{cslck1}), $j=N+1$ is not included since this constraint is in
fact satisfied with equality, because otherwise the objective function
can be increased by increasing some $p_i$. Note also that as $p_i>0$,
we did not include any slackness conditions for $p_i$.

We apply the KKT optimality conditions to this Lagrangian to obtain
the optimal power levels $p_i^*$ in terms of the Lagrange multipliers
as,
\begin{align} 
\label{sol1} 
p_i^* = \frac{1}{\left(\sum_{j=i}^{N+1} \lambda_j - \sum_{j=i}^{N} \mu_j  \right)}-1, \quad i=1,\ldots,N
\end{align} 
and $p^*_{N+1}= \frac{1}{\lambda_{N+1}}- 1$. Note that $p_i^*$ that satisfy $\sum_{i=1}^{N+1} L_ip_i^* = \sum_{i=0}^{N}E_i$ is unique.

Based on the expression for $p_i^*$ in terms of the Lagrange multipliers in (\ref{sol1}), we have the following observation on the structure of the optimal power allocation scheme.
\begin{theorem}
\label{first}
When $E_{max}=\infty$, the optimal power levels is a monotonically increasing sequence: $p_{i+1}^* \geq p_i^*$. Moreover, if for some $\ell$, $\sum_{i=1}^{\ell} L_i p_i^* < \sum_{i=0}^{\ell-1} E_i$, then $p_{\ell}^* = p_{\ell+1}^*$.
\end{theorem}
\begin{Proof}
Since $E_{max}=\infty$, constraints in (\ref{opt12}) are satisfied without equality and $\mu_i = 0$ for all $i$ by slackness conditions in (\ref{cslck2}). From (\ref{sol1}), since $\lambda_i \geq 0$, optimum $p_i^*$ are monotonically increasing: $p_{i+1}^* \geq p_i^*$. Moreover, if for some $\ell$, $\sum_{i=1}^{\ell} L_i p_i^* < \sum_{i=0}^{\ell-1} E_i$, then $\lambda_\ell = 0$, which means $p_{\ell}^* = p_{\ell+1}^*$.
\end{Proof}

The monotonicity in Theorem~\ref{first} is a result of the fact that energy may be spread from the current time to the future for optimal operation. Whenever a constraint in (\ref{opt11}) is not satisfied with equality, it means that some energy is available for use but is not used in the current epoch and is transferred to future epochs. Hence, the optimal power allocation is such that, if some energy is transferred to future epochs, then the power level must remain the same. However, if the optimal power level changes from epoch $i$ to $i+1$, then this change should be in the form of an increase and no energy is transferred for future use. That is, the corresponding constraint in (\ref{opt11}) is satisfied with equality.

If $E_{max}$ is finite, then its effect on the optimal power allocation is observed through $\mu_i$ in (\ref{sol1}). In particular, if the constraints in (\ref{opt12}) are satisfied without equality, then optimal $p_i^*$ are still monotonically increasing since $\mu_i=0$. However, as $E_i \leq E_{max}$ for all $i$, the constraint with the same index in (\ref{opt11}) is satisfied without equality whenever a constraint in (\ref{opt12}) is satisfied with equality. Therefore, a non-zero $\mu_i$ and a zero $\lambda_i$ appear in $p_i^*$ in (\ref{sol1}). This implies that the monotonicity of $p_i^*$ may no longer hold. $E_{max}$ constraint restricts  power levels to take the same value in adjacent epochs as it constrains the energy that can be transferred from current epoch to the future epochs. Indeed, from constraints in (\ref{opt12}), the energy that can be transferred from current, say the $i$th, or previous epochs, to future epochs is $E_{max} - E_i$. Hence, the power levels are equalized only to the extent that $E_{max}$ constraint allows. 

\subsection{Directional Water-Filling Algorithm}
We interpret the observed properties of the optimal power allocation scheme as a {\it directional water-filling} scheme. We note that if $E$ units of water (energy) is filled into a rectangle of bottom size $L$, then the water level is $\frac{E}{L}$. Another key ingredient of the directional water-filling algorithm is the concept of a {\it right permeable} tap, which permits transfer of water (energy) only from left to right. 

\begin{figure}[t]
\centering
\subfigure{
\includegraphics[width=0.7\linewidth]{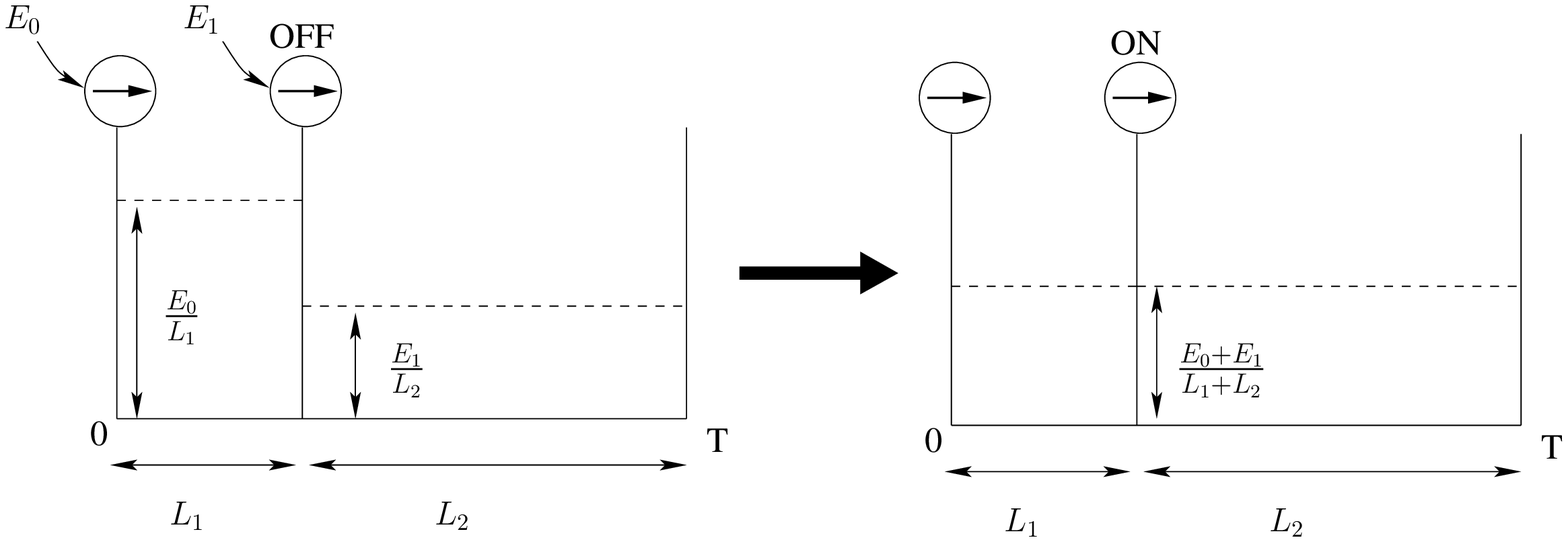}
\label{xx}
}
\subfigure{
\includegraphics[width=0.7\linewidth]{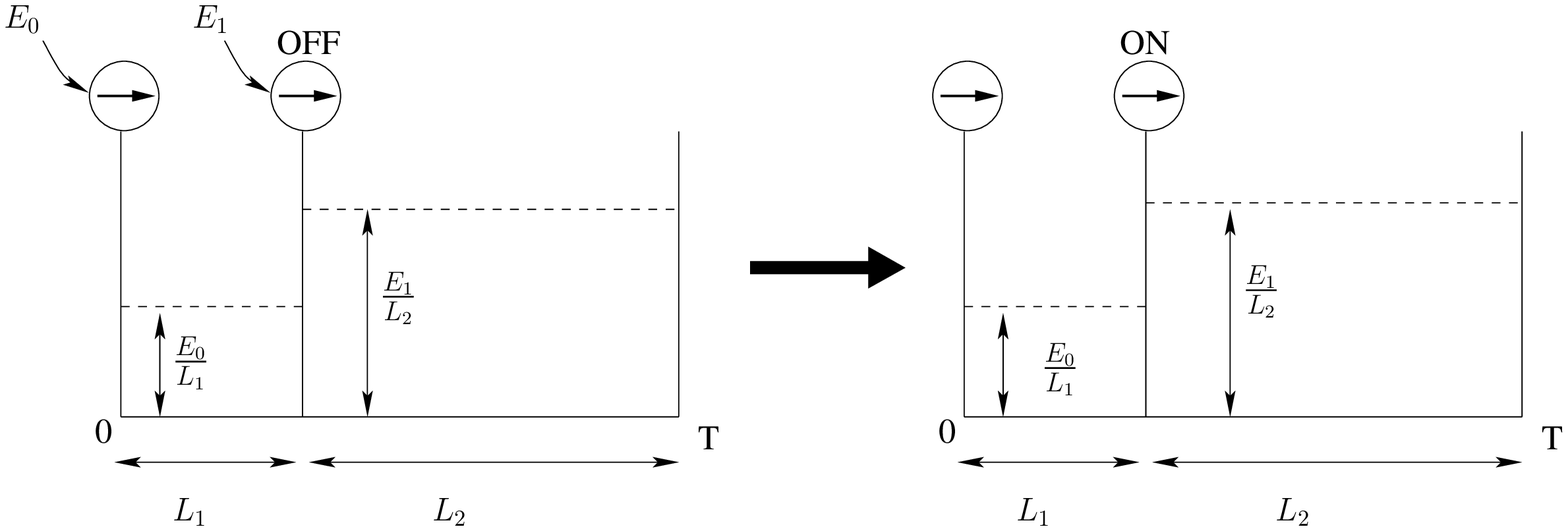}
\label{xy1}
}
\subfigure{
\includegraphics[width=0.7\linewidth]{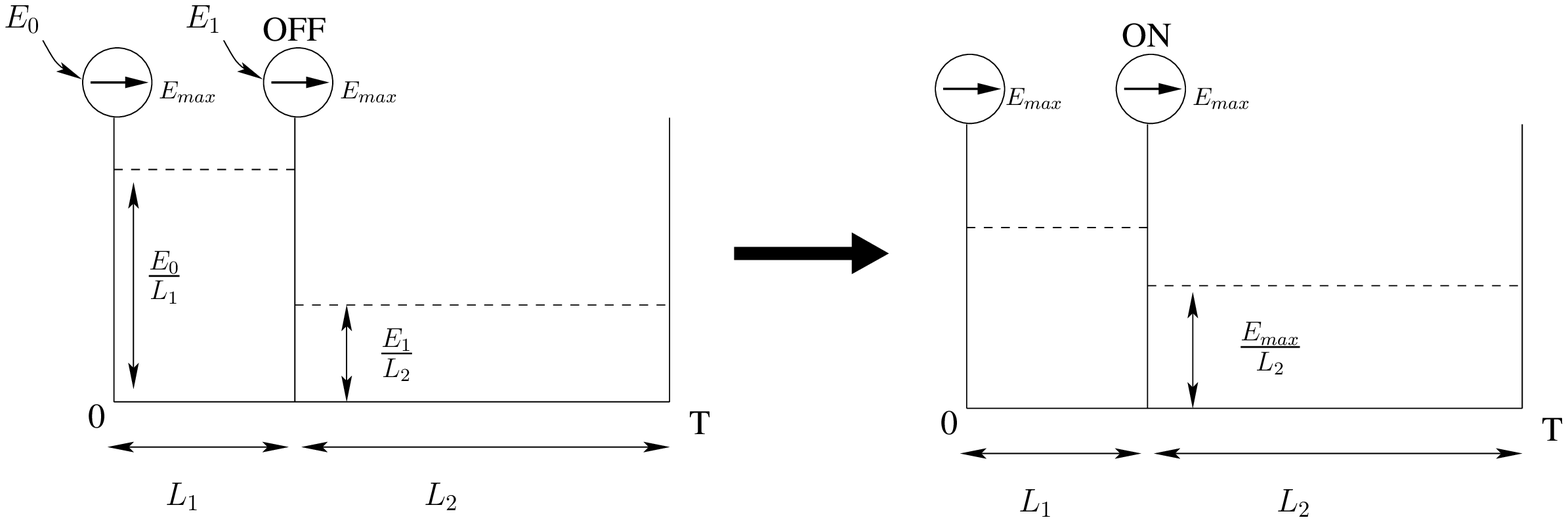}
\label{xy2}
}
\caption{Directional water-filling with right permeable taps in a two-epoch setting.}
\label{twoepo1}
\end{figure}

Consider the two epoch system. Assume $E_{max}$ is sufficiently large. If $\frac{E_0}{L_1} > \frac{E_1}{L_2}$, then some energy is transferred from epoch $1$ to epoch $2$ so that the levels are equalized. This is shown in the top figure in Fig.~\ref{twoepo1}. However, if $\frac{E_0}{L_1} < \frac{E_1}{L_2}$, no energy can flow from right to left. This is due to the causality of energy usage, i.e., energy cannot be used before it is harvested. Therefore, as shown in the middle figure in Fig.~\ref{twoepo1}, the water levels are not equalized. We implement this using {\it right permeable} taps, which let water (energy) flow only from left to right.

We note that the finite $E_{max}$ case can be incorporated into the energy-water analogy as a constraint on the amount of energy that can be transferred from the past to the future. If the equalizing water level requires more than $E_{max}-E_i$ amount of energy to be transferred, then only $E_{max} - E_i$ can be transferred. Because, otherwise, the energy level in the next epoch exceeds $E_{max}$ causing overflow of energy, which results in inefficiencies. More specifically, when the right permeable tap in between the two epochs of the example in bottom figure in Fig.~\ref{twoepo1} is turned on, only $E_{max} - E_1$ amount of energy transfer is allowed from epoch $1$ to epoch $2$. 

\section{Maximizing Throughput in a Fading Channel}
We now solve for the offline policy for the fading channel utilizing the insights obtained in the previous section. The channel state changes $M$ times and energy arrives $N$ times in the duration $[0, T)$. Hence, we have $M+N+1$ epochs. Our goal is again to maximize the number of bits transmitted by the deadline $T$. Similar to the non-fading case, the optimal power management strategy is such that the transmit power is constant in each event epoch. Therefore, let us again denote the transmit power in epoch $i$ by $p_i$, for $i=1,\ldots,M+N+1$. We define $E_{in}(i)$ as the energy which arrives in epoch $i$. Hence, $E_{in}(i)=E_j$ for some $j$ if event $i$ is an energy arrival and $E_{in}(i)=0$ if event $i$ is a fade level change. Also, $E_{in}(1)=E_0$. Similar to the non-fading case, we have causality constraints due to energy arrivals and an $E_{max}$ constraint due to finite battery size. Hence, the optimization problem in this fading case becomes:
\begin{eqnarray}
\label{fdngop}
\max_{p_i\geq 0} &  & \sum_{i=1}^{M+N+1} \frac{L_i}{2}\log\left(1 + h_i p_i\right)  \\ \label{refc}
\mbox{s.t.} & & \sum_{i=1}^{\ell} L_i p_i \leq \sum_{i=1}^\ell E_{in}(i), \quad \forall\ell \\
 & & \sum_{i=1}^\ell E_{in}(i) - \sum_{i=1}^\ell L_i p_i \leq E_{max}, \quad \forall \ell
\end{eqnarray}
Note that, as in the non-fading case, the constraint in (\ref{refc}) for $\ell = M+N+1$ must be satisfied with equality, since otherwise, we can always increase one of $p_i$ to increase the throughput.

As in the non-fading case, the objective function in (\ref{fdngop}) is concave and the constraints are convex. The optimization problem has a unique optimal solution. We define the Lagrangian for any $\lambda_i$, $\mu_i$ and $\eta_i$ as,
\begin{align} 
\label{lgrn} \nonumber 
\mathcal{L} = & \sum_{i=1}^{M+N+1} \frac{L_i}{2}\log\left(1 + h_i p_i\right) - \sum_{j=1}^{M+N+1} \lambda_j \left( \sum_{i=1}^j L_ip_i - \sum_{i=1}^{j}E_{in}(i)  \right) \\ &- \sum_{j=1}^{M+N+1} \mu_j\left(\sum_{i=1}^j E_{in}(i) - \sum_{i=1}^j L_ip_i - E_{max}\right) + \sum_{i=1}^{M+N+1} \eta_i p_i  
\end{align}
Note that we have not employed the Lagrange multipliers $\{\eta_i\}$ in the non-fading case, since in that case, we need to have all $p_i>0$. However, in the fading case, some of the optimal powers can be zero depending on the channel fading state. Associated complimentary slackness conditions are,
\begin{align}
\label{slc1}
\lambda_j \left(\sum_{i=1}^j L_ip_i - \sum_{i=1}^{j}E_{in}(i)\right) & = 0, \quad \forall j
\\ \label{slc2}
\mu_j\left(\sum_{i=1}^j E_{in}(i) - \sum_{i=1}^j L_ip_i - E_{max}\right) & = 0, \quad \forall j 
\\ \label{slc3}
\eta_jp_j & = 0, \quad \forall j
\end{align}
It follows that the optimal powers are given by
\begin{align}
p_i^* = \left[ \nu_i - \frac{1}{h_i} \right]^+
\label{pow-sol}
\end{align}
where the water level in epoch $i$, $\nu_i$, is
\begin{align} 
\label{water} 
\nu_i= \frac{1}{\sum_{j=i}^{M+N+1}\lambda_j - \sum_{j=i}^{M+N+1}\mu_j}
\end{align}
We have the following observation for the fading case.
\begin{theorem}
\label{res1}
When $E_{max}=\infty$, for any epoch $i$, the optimum water level $\nu_i$ is monotonically increasing: $\nu_{i+1}\geq \nu_i$. Moreover, if some energy is transferred from epoch $i$ to $i+1$, then $\nu_i=\nu_{i+1}$.  
\end{theorem}
\begin{Proof}
$E_{max}=\infty$ assumption results in $\mu_i=0$ for all $i$. From (\ref{water}), and since $\lambda_i \geq 0$, we have $\nu_{i+1} \geq \nu_i$. If energy is transferred from the $i$th epoch to the $i+1$st epoch, then the $i$th constraint in (\ref{refc}) is satisfied without equality. This implies, by slackness conditions in (\ref{slc1}), that for those $i$, we have $\lambda_i=0$. Hence, by (\ref{water}), $\nu_i = \nu_{i+1}$. In particular, $\nu_i = \nu_j$ for all epochs $i$ and $j$ that are in between two consecutive energy arrivals as there is no wall between these epochs and injected energy freely spreads into these epochs.
\end{Proof}

As in the non-fading case, the effect of finite $E_{max}$ is observed via the Lagrange multipliers $\mu_i$. In particular, whenever $E_{max}$ constraint is satisfied with equality, the monotonicity of the water level no longer holds. $E_{max}$ constrains the amount of energy that can be transferred from one epoch to the next. Specifically, the transferred energy cannot be larger than $E_{max} - E_{in}(i)$. Note that this constraint is trivially satisfied for those epochs with $E_{in}(i)=0$ because $E_{in}(i)< E_{max}$ and hence the water level in between two energy arrivals must be equalized. However, the next water level may be higher or lower depending on the new arriving energy amount.

\subsection{Directional Water-Filling Algorithm}
\begin{figure}[t]
\begin{center}
\includegraphics[width=0.7\linewidth]{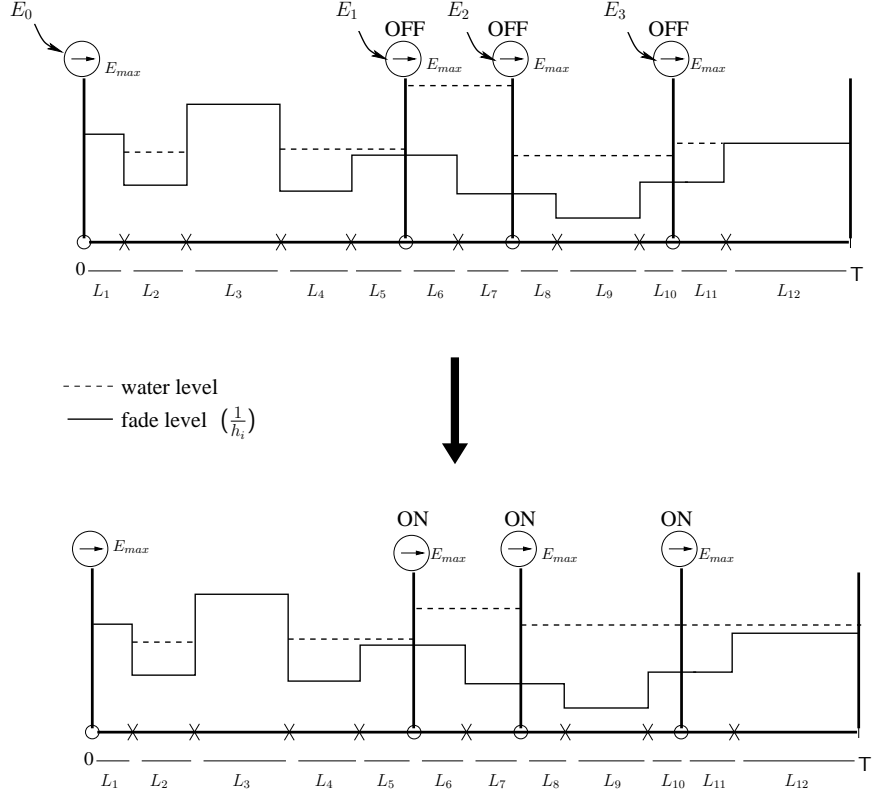}
\end{center}
\caption{Directional water-filling with right permeable taps in a fading channel.}
\label{fadesm}
\end{figure}

The directional water-filling algorithm in the fading channel requires walls at the points of energy arrival, with right permeable water taps in each wall which allows at most $E_{max}$ amount of water to flow. No walls are required to separate the epochs due to changes in the fading level. The water levels when each right permeable tap is turned on will be found by the directional water-filling algorithm. Optimal power allocation $p_i^*$ is then calculated by plugging the resulting water levels into (\ref{pow-sol}). An example run of the algorithm is shown in Fig.~\ref{fadesm}, for a case of 12 epochs. Three energy arrivals occur during the course of the transmission, in addition to the energy available at time $t=0$. We observe that the energy level equalizes in epochs 2, 4, 5, while no power is transmitted in epochs 1 and 3, since the channel gains in these epochs are too low (i.e., $\frac{1}{h_i}$ too high). The energy arriving at the beginning of epoch 6 cannot flow left due to causality constraints, which are enforced by right permeable taps, which allow energy flow only to the right. We observe that the energy equalizes between epochs 8 through 12, however, the excess energy in epochs 6 and 7 cannot flow right, due to the $E_{max}$ constraint enforced by the right permeable tap between epochs 7 and 8. 

\section{Transmission Completion Time Minimization in Fading Channel}
In contrast to the infinite backlog assumption of the previous sections, we now assume that the transmitter has $B$ bits to be communicated to the receiver in the energy harvesting and fading channel setting. Our objective now is to minimize the time necessary to transmit these $B$ bits. This problem is called the transmission completion time minimization problem. In \cite{jing10ciss, tcom-submit}, this problem is formulated and solved for an energy harvesting system in a non-fading environment. In \cite{kaya_subm}, the problem is solved when there is an $E_{max}$ constraint on the energy buffer (battery) by identifying its connection to its throughput optimization counterpart. Here, our goal is to address this problem in a fading channel, by using the directional water-filling approach we have developed so far.

The transmission completion time minimization problem can be stated as,
\begin{eqnarray}
\min & & T  \label{tctm} \\ 
\mbox{s.t.} & & \sum_{i=1}^{N} \frac{L_i}{2}\log\left(1 + h_i p_i\right)=B \label{tctm-con1} \\
& & \sum_{i=1}^{\ell} L_i p_i \leq \sum_{i=1}^\ell E_{in}(i), \quad \ell=1,\ldots, N \label{tctm-con2} \\ 
& & \sum_{i=1}^\ell E_{in}(i) - \sum_{i=1}^{\ell} L_i p_i \leq E_{max}, \quad \ell=1,\ldots,N~ \label{tctm-con3} \quad
\end{eqnarray}
where $N \triangleq N(T)$ is the number of epochs in the interval $[0,T]$. The solution will be a generalization of the results in \cite{jing10ciss, tcom-submit, kaya_subm} for the fading case. To this end, we introduce the maximum departure curve. This maximum departure curve function will map the transmission completion time minimization problem of this section to the throughput maximization problem of the previous sections.

\subsection{Maximum Departure Curve}
Given a deadline $T$, define the maximum departure curve $D(T)$ for a given sequence of energy arrivals and channel fading states as, 
\begin{align} 
\label{mx} 
D(T) = \max \sum_{i=1}^{N(T)} \frac{L_i}{2}\log\left(1 + h_ip_i \right) 
\end{align} 
where $N(T)$ is the number of epochs in the interval $[0,T]$. The maximization in (\ref{mx}) is subject to the energy causality and maximum battery storage capacity constraints in (\ref{tctm-con2}) and (\ref{tctm-con3}). The maximum departure function $D(T)$ represents the maximum number of bits that can be served out of the backlog by the deadline $T$ given the energy arrival and fading sequences. This is exactly the solution of the problem studied in the previous sections. Some characteristics of the maximum departure curve are stated in the following lemma.
\begin{lemma}
\label{cont}
The maximum departure curve $D(T)$ is a monotonically increasing and continuous function of $T$. $D(T)$ is not differentiable at $\{t_i^e\}$ and $\{t_i^f\}$. 
\end{lemma}
\begin{Proof}
The monotonicity follows because as the deadline is increased, we can transmit at least as many bits as we could with the smaller deadline. The continuity follows by observing that, if no new energy arrives or fading state changes, there is no reason to have a discontinuity. When new energy arrives, since the number of bits that can be transmitted with a finite amount of energy is finite, the number of bits transmitted will not have any jumps. Similarly, if the fading level changes, due to the continuity of the $\log$ function, $D(T)$ will be continuous.

For the non-differentiable points, assume that at $t=t_i^e$, an energy in the amount of $E_i$ arrives. There exists a small enough increment from $t_i^e$ that the water level on the right is higher than the water level on the left. The right permeable taps will not allow this water to flow to left. Then, the $D(T)$ will have the following form:
\begin{align} 
D(t_i^e + \Delta) = D(t_i^e) + \frac{\Delta}{2}\log\left(1+\frac{E_i h}{\Delta}\right) \end{align} 
Thus, the right derivative of $D(T)$ at $t=t_i^e$, becomes arbitrarily large. Hence, $D(T)$ is not differentiable at $t_i^e$. 
At $t=t_i^f$, the fade level changes from $h_i$ to $h_{i+1}$. As $t$ is increased, water level decreases unless new energy arrives. The change in the water level is proportional to $\frac{1}{h_{i+1}}$ for $t>t_i^f$ and is proportional to $\frac{1}{h_i}$ for $t<t_i^f$. Hence, at $t=t_i^f$, $D(T)$ is not differentiable. 
\end{Proof}

The continuity and monotonicity of $D(T)$ implies that the inverse function of $D(T)$ exists, and that for a closed interval $[a,b]$, $D^{-1}([a,b])$ is also a closed interval. Since $D(T)$ is obtained by the directional water-filling algorithm, the derivative of $D(T)$ has the interpretation of the rate of energy transfer from past into the future at time $T$, i.e., it is the measure of the tendency of water to flow right. The non-differentiabilities at energy arrival and fading change points are compatible with this interpretation. 

We can visualize the result of Lemma~\ref{cont} by considering a few simple examples. As the simplest example, consider the non-fading channel ($h=1$) with $E_0$ units of energy available at the transmitter (i.e., no energy arrivals). Then, the optimal transmission scheme is a constant transmit power scheme, and hence, we have, \begin{align}
D(T)=\frac{T}{2}\log\left(1 + \frac{E_0}{T}\right)
\end{align}
It is clear that this is a continuous, monotonically increasing function, whose derivative at $T=0$ (at the time of energy arrival) is unbounded.

Next, we consider a two epoch case where $E_1$ arrives at $T_1$ and fading level is constant (and also $h=1$). We assume $E_0$ and $E_1$ are both smaller than $E_{max}$ and $E_0 + E_1 > E_{max}$. After some algebra, $D(T)$ can shown to be expressed as, 
\begin{align}
\label{msc22}
D(t) =\left\{\begin{array}{ll}
\frac{t}{2}\log\left(1 + \frac{E_0}{t}\right),  & 0 < t < T_1\\
\frac{T_1}{2}\log\left(1 + \frac{E_0}{T_1}\right) + \frac{t-T_1}{2}\log\left(1 + \frac{E_1}{t-T_1}\right), & T_1\leq t \leq T_2\\
\frac{t}{2}\log\left(1 + \frac{E_0+E_1}{t}\right),  & T_2<t<T_3\\
\frac{T_3}{2}\log\left(1 + \frac{E_0+E_1-E_{max}}{T_3}\right) + \frac{t-T_3}{2}\log\left(1 + \frac{E_{max}}{t-T_3}\right), & T_3<t<\infty
\end{array}
\right. 
\end{align}
where $T_2=\frac{E_1T_1}{E_0} + T_1$, $T_3=\frac{T_1(E_0 + E_1)}{E_0+E_1-E_{max}}$. In this $E_{max}$ constrained case, the asymptote of $D(T)$ as $T\rightarrow \infty$ is strictly smaller than that in $E_{max}=\infty$ case. 

\begin{figure}[t]
\begin{center}
\includegraphics[width=0.4\linewidth]{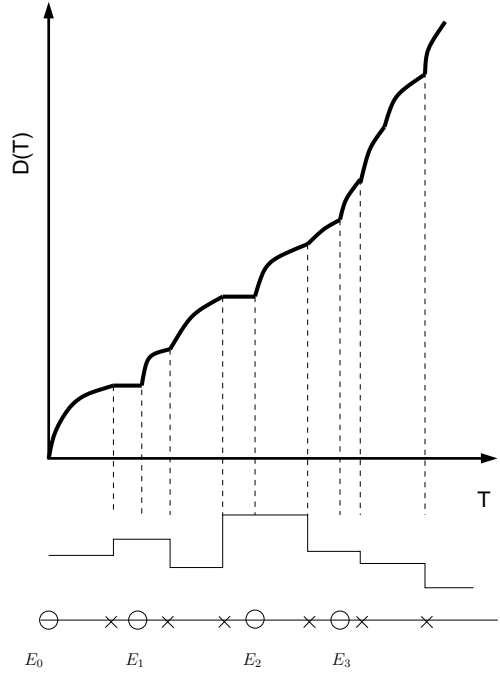}
\end{center}
\caption{The general form of the maximum departure curve.}
\label{maxserv4}
\end{figure}

In the most general case where we have multiple energy arrivals and channel state changes, these basic properties will follow. An example case is shown in Fig.~\ref{maxserv4}. Note that there may be discontinuities in $D'(T)$ due to other reasons than fading level changes and energy arrivals, such as the $E_{max}$ constraint. 

\subsection{Solution of the Transmission Completion Time Minimization Problem in a Fading Channel}
We now solve the transmission completion time minimization problem stated in (\ref{tctm})-(\ref{tctm-con3}). Minimization of the time to complete the transmission of $B$ bits available at the transmitter is closely related with the maximization of the number of bits that can be sent by a deadline. In fact, if the maximum number of bits that can be sent by $T$ is less than $B$, then it is not possible to complete the transmission of $B$ bits by $T$. As we state formally below, if $T^*$ is the minimal time to complete the transmission of $B$ bits, then necessarily $B=D(T^*)$. This argument provides a characterization for $T^*$ in terms of the maximum departure curve, as stated in the following theorem.

\begin{theorem}
The minimum transmission completion time $T^*$ to transmit $B$ bits is $T^* = \min\{t \in \mathcal{M}_{B}\}$ where $\mathcal{M}_{B} = \{t: B = D(t)\}$.
\end{theorem}
\begin{Proof}
For $t$ such that $D(t)<B$, $T^*>t$ since the maximum number of bits that can be served by $t$ is $D(t)$ and it is less than $B$. Hence, $B$ bits cannot be completed by $t$. Conversely, for $t$ such that $D(t)>B$, $T^*<t$ because $B$ bits can be completed by $t$. Hence, $D(T^*)=B$ is a necessary condition. As $D(T)$ is continuous, the set $\{t: B = D(t)\}$ is a closed set. Hence, $\min\{ t: B = D(t)\}$ exists and is unique. By the definition of $T^*$, we have $T^*=\min\{t: B = D(t)\}$. 
\end{Proof}

\section{Online Transmission Policies}

In this section, we will study scheduling in the given setting with online, i.e., causal, information of the events. In particular, we consider the maximization of the number of bits sent by deadline $T$ given only causal information of the energy arrivals and channel fade levels at the transmitter side as in Fig. \ref{mod1}.

We assume that the energy arrival is a compound Poisson process with a density function $f_e$. Hence, $N_e$ is a Poisson random variable with mean $\lambda_eT$. The channel fade level is a stochastic process marked with a Poisson process of rate $\lambda_f$. Thus, $N_f$ is Poisson with mean $\lambda_fT$. The channel takes independent values with probability density $f_h$ at each marked time and remains constant in between two marked points.  

\subsection{Optimal Online Policy}
The states of the system are fade level $h$ and battery energy $e$. An online policy is denoted as $g(e,h,t)$ which denotes the transmit power decided by the transmitter at time $t$ given the states $e$ and $h$. We call a policy admissible if $g$ is nonnegative, $g(0,h,t)=0$ for all $h$ and $t\in[0,T]$ and $e(T)=0$. That is, we impose an infinite cost if the remaining energy in the battery is non-zero after the deadline. Hence, admissible policies guarantee that no transmission can occur if the battery energy is zero and energy left in the battery at the time of the deadline is zero so that resources are used fully by the deadline. The throughput $J_g(e,h,t)$ is the expected number of bits sent by the time $t$ under the policy $g$
\begin{align} 
\label{value1} 
J_g(e,h,t)= E\left[ \int_0^t \frac{1}{2}\log\left(1 + h(\tau)g(e,h,\tau)\right)d\tau\right] 
\end{align}
Then, the value function is the supremum over all admissible policies $g$ 
\begin{align} 
\label{value2} 
J(e,h,t)= \sup_g J_g 
\end{align} 
Therefore, the optimal online policy $g^*(e,h,t)$ is such that $J(e,h,t=0)=J_{g^*}$, i.e., it solves the following problem 
\begin{align} 
\label{onl_mx} 
\max_{g} E\left[ \int_0^T \frac{1}{2}\log\left(1 + h(\tau)g(e,h,\tau)\right)d\tau\right] 
\end{align} 
In order to solve (\ref{onl_mx}), we first consider $\delta$-skeleton of the random processes \cite{zafer05alerton}. For sufficiently small $\delta$, we quantize the time by $\delta$ and have the following. \begin{align}\nonumber &\max_{g} E\left[ \int_0^T \frac{1}{2}\log\left(1 + h(\tau)g(e,h,\tau)\right)d\tau\right] = \\ \label{lb} & \max_{g(e,h,0)} \left( \frac{\delta}{2}\log\left(1 + h(0)g(e,h,0)\right) + J(e-\delta g(e,h,0),h,\delta) \right) \end{align}
Then, we can recursively solve (\ref{lb}) to obtain $g^*(e,h,t=k\delta)$ for $k=1,2,\ldots,\lfloor{\frac{T}{\delta}\rfloor}$. This procedure is the dynamic programming solution for continuous time and the outcome is the optimal online policy \cite{zafer05alerton,BersekasDynProg}. After solving for $g^*(e,h,t)$, the transmitter records this function as a look-up table and at each time $t$, it receives feedback $h(t)$, senses the battery energy $e(t)$ and transmits with power $g^*(e(t),h(t),t)$. 

\subsection{Other Online Policies}

Due to the {\it curse of dimensionality} inherent in the dynamic programming solution, it is natural to forgo performance in lieu of less complex online policies. In this subsection, we propose several suboptimal transmission policies that can somewhat mimic the offline optimal algorithms while being computationally simpler and requiring less statistical knowledge. In particular, we resort to event-based online policies which react to a change in fading level or an energy arrival. Whenever an event is detected, the online policy decides on a new power level. Note that the transmission is subject to availability of energy and the $E_{max}$ constraint.

\subsubsection{Constant Water Level Policy}
The constant water level policy makes online decisions for the transmit power whenever a change in fading level is observed through the causal feedback. Assuming that the knowledge of the average recharge rate $P$ is available to the transmitter and that fading density $f_h$ is known, the policy calculates $h_0$ that solves the following equation. \begin{align} \int_{h_0}^\infty \left(\frac{1}{h_0} - \frac{1}{h}\right)f_h(h) dh = P \end{align} Whenever a change in the fading level occurs, the policy decides on the following power level $p_i = \left( \frac{1}{h_0} - \frac{1}{h_i} \right)^+$. If the energy in the battery is nonzero, transmission with $p_i$ is allowed, otherwise the transmitter becomes silent.

Note that this power control policy is the same as the capacity achieving power control policy in a stationary fading channel \cite{Goldsmith97} with an average power constraint equal to the average recharge rate. In~\cite{sharma10TWC}, this policy is proved to be stability optimal in the sense that all data queues with stabilizable arrival rates can be stabilized by policies in this form where the power budget is $P-\epsilon$ for some $\epsilon>0$ sufficiently small. However, for the time constrained setting, this policy is strictly suboptimal as will be verified in the numerical results section. This policy requires the transmitter to know the mean value of the energy arrival process and the full statistics of the channel fading. A channel state information (CSI) feedback is required from the receiver to the transmitter at the times of events only.

\subsubsection{Energy Adaptive Water-Filling}
Another reduced complexity event-based policy is obtained by adapting the water level to the energy level in each event. Again the fading statistics is assumed to be known. Whenever an event occurs, the policy determines a new power level. In particular, the cutoff fade level $h_0$ is calculated at each energy arrival time as the solution of the following equation \begin{align} \int_{h_0}^\infty \left(\frac{1}{h_0} - \frac{1}{h}\right)f(h) dh = E_{current} \end{align} where $E_{current}$ is the energy level at the time of the event. Then, the transmission power level is determined similarly as $p_i = \left( \frac{1}{h_0} - \frac{1}{h} \right)^+$. This policy requires transmitter to know the fading statistics. A CSI feedback is required from the receiver to the transmitter at the times of changes in the channel state.

\subsubsection{Time-Energy Adaptive Water-Filling}
A variant of the energy adaptive water filling policy is obtained by adapting the power to the energy level and the remaining time to the deadline. The cutoff fade level $h_0$ is calculated at each energy arrival time as the solution of the following equation. \begin{align} \int_{h_0}^\infty \left(\frac{1}{h_0} - \frac{1}{h}\right)f(h) dh = \frac{E_{current}}{T-s_i} \end{align} Then, the transmission power level is determined as $p_i = \left( \frac{1}{h_0} - \frac{1}{h} \right)^+$. 

\section{Numerical Results} 
\label{sect_numerical}

We consider a fading additive Gaussian channel with bandwidth $W$ where the instantaneous rate is \begin{align} r(t) = W\log\left(1 + h(t)p(t)\right) \end{align} $h(t)$ is the channel SNR, i.e., the actual channel gain divided by the noise power spectral density multiplied by the bandwidth, and $p(t)$ is the transmit power at time $t$. Bandwidth is chosen as $W=1$ MHz for the simulations.

We will examine the deadline constrained throughput performances of the optimal offline policy, optimal online policy, and other proposed sub-optimal online policies. In particular, we compare the optimal performance with the proposed sub-optimal online policies which are based on waterfilling \cite{Goldsmith97}. The proposed sub-optimal online policies use the fading distribution, and react only to the new energy arrivals and fading level changes. These event-based algorithms require less feedback and less computation, however, the fact that they react only to the changes in the fading level and new energy arrivals is a shortcoming of these policies. Since the system is deadline constrained, the policies need to take the remaining time into account yet the proposed policies do not do this optimally. We will simulate these policies under various different settings and we will observe that the proposed sub-optimal policies may perform very well in some cases while not as well in some others.   
      
We perform all simulations for 1000 randomly generated realizations of the channel fade pattern and $\delta=0.001$ is taken for the calculation of the optimal online policy. The rates of Poisson mark processes for energy arrival and channel fading $\lambda_e$ and $\lambda_f$ are assumed to be 1. The unit of $\lambda_e$ is $\text{J}/\text{sec}$ and that of $\lambda_f$ is ${1}/{\text{sec}}$. Hence, the mean value of the density function $f_e$ is also the average recharge rate and the mean value of $f_h$ is the average fading level. The changes in the fading level occur relatively slowly with respect to the symbol duration. 

$f_e$ is set as a non-negative uniform random variable with mean $P$, and as the energy arrival is assumed to be smaller than $E_{max}$, we have $2P<E_{max}$. Selection of the $E_{max}$ constraint is just for illustration. In real life, sensors may have batteries of $E_{max}$ on the order of $\text{kJ}$ but the battery feeds all circuits in the system. Here, we assume a fictitious battery that carries energy for only communication purposes. Hence, $E_{max}$ on the order of $1$ J will be considered. We will examine different fading distributions $f_h$. In particular, Nakagami distribution with different shape parameter $m$ will be considered. We implement the specified fading by sampling its probability density function with sufficiently large number of points. 

In order to assess the performance, we find an upper bound on the performances of the policies by first assuming that the channel fading levels and energy arrivals in the $[0,T]$ interval are known non-causally, and that the total energy that will arrive in $[0,T]$ is available at the transmitter at time $t=0$. Then, for the water level $p_w$ that is obtained by spreading the total energy to the interval $[0,T]$, with the corresponding fading levels, yield the throughput $T^{ub}$ defined in the following  
\begin{align} 
T^{ub} = \frac{W}{T} \sum_{i=1}^K l_i \frac{1}{2}\log\left( 1 + h_i\left( p_w - \frac{1}{h_i} \right)^+\right) 
\end{align}
as an upper bound for the average throughput in the $[0,T]$ interval; here $l_i$ denotes the duration of the fade level in the $i$th epoch.
Even the offline optimal policy has a smaller average throughput than $T^{ub}$ as the causality constraint does not allow energies to be spread evenly into the entire interval.

\begin{figure}[t]
\begin{center}
\includegraphics[width=0.65\linewidth]{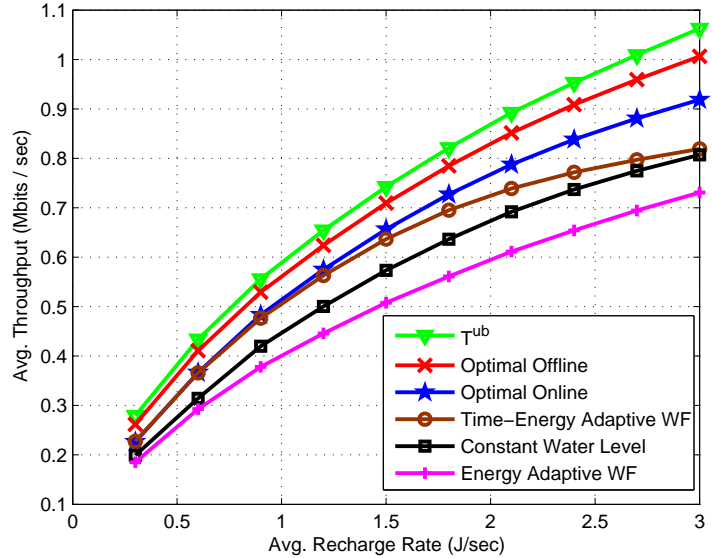}
\end{center}
\caption{Performances of the policies for various energy arrival rates under unit-mean Rayleigh fading, $T=10$ sec and $E_{max}=10$ J.}
\label{sim1}
\end{figure}

\begin{figure}[t]
\begin{center}
\includegraphics[width=0.65\linewidth]{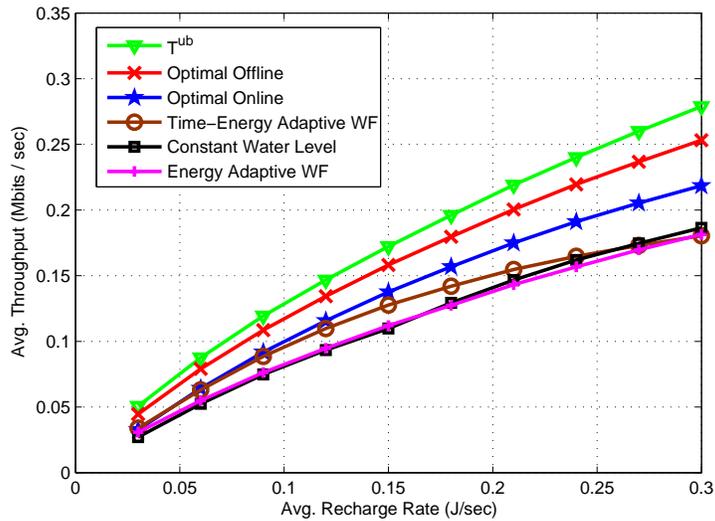}
\end{center}
\caption{Performances of the policies for various average recharge rates under unit-mean Rayleigh fading, $T=10$ sec and $E_{max}=1$ J.}
\label{sim2}
\end{figure}

We start with examining the average throughput of the system under Rayleigh fading with SNR$=0$ dB and deadline $T=10$ sec, $E_{max}=10$ J as depicted in Fig. \ref{sim1}. We observe that time-energy adaptive waterfilling policy performs quite close to the optimal online policy in the low recharge rate regime. It can be a viable policy to spread the incoming energy when the recharge rate is low; however, its performance saturates as the recharge rate is increased. In this case the incoming energy cannot be easily accommodated and more and more energy is lost due to overflows. Similar trends can be found in Fig. \ref{sim2} under very low recharge rate regime in the same setting with only difference being the battery capacity $E_{max}=1$ J. Next, we examine the setting with $T=10$ sec, $E_{max}=10$ J under Nakagami fading of $m=3$ (average SNR$=5$ dB) and we observe similar performances as in the previous cases. As a common behavior in these settings, energy adaptive water-filling performs poorer with respect to the constant water level and time-energy adaptive water-filling schemes.  

\begin{figure}[t]
\begin{center}
\includegraphics[width=0.65\linewidth]{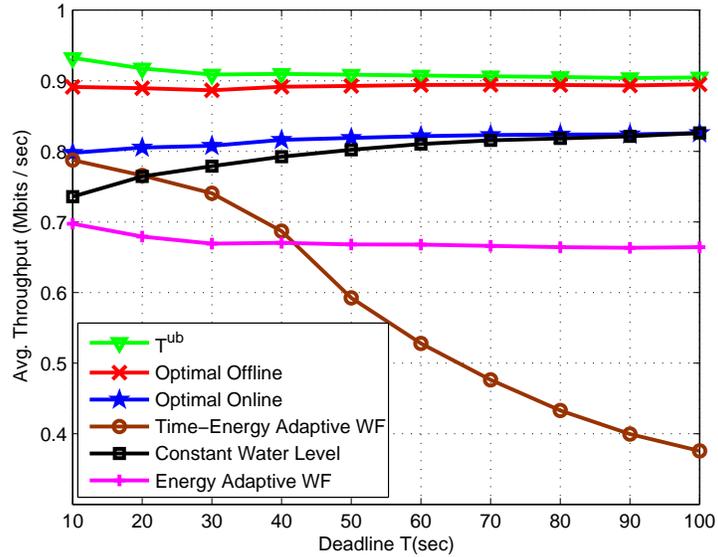}
\end{center}
\caption{Performances of the policies with respect to deadline $T$ under Nakagami fading distribution with $m=5$ and average recharge rate $P=0.5$ J/sec and $E_{max}=10$ J.}
\label{sim6}
\end{figure}

Finally, we examine the policies under different deadline constraints and present the plots for Nakagami fading distribution with $m=5$ in Fig. \ref{sim6}. A remarkable result is that as the deadline is increased, stability optimal \cite{sharma10TWC} constant water level policy approaches the optimal online policy. We conclude that the time-awareness of the optimal online policy has less and less importance as the deadline constraint becomes looser. We also observe that the throughput of the energy-adaptive waterfilling policy is roughly a constant regardless of the deadline. Moreover, the time-energy adaptive policy performs worse as $T$ is increased because energies are spread to very long intervals rendering the transmit power very small and hence energy accumulates in the battery. This leads to significant energy overflows since the battery capacity is limited, and the performance degrades. 

\begin{figure}[t]
\begin{center}
\includegraphics[width=0.65\linewidth]{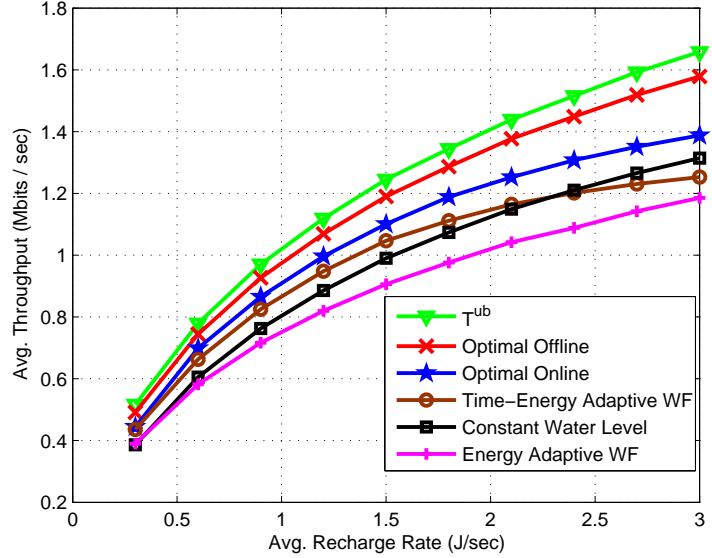}
\end{center}
\caption{Performances of the policies for different energy recharge rates under Nakagami fading with $m=3$, $T=10$ sec and $E_{max}=10$ J.}
\label{sim5}
\end{figure}

\section{Conclusions}
We developed optimal energy management schemes for energy harvesting
systems operating in fading channels, with finite capacity
rechargeable batteries. We considered two related problems under offline knowledge of the events: maximizing
the number of bits sent by a deadline, and minimizing the time it
takes to send a given amount of data. We solved the first problem
using a directional water-filling approach. We solved the second
problem by mapping it to the first problem via the maximum departure
curve function. Finally, we solved for throughput optimal policy for the deadline constrained setting under online knowledge of the events using dynamic programming in continuous time. Our numerical results show the performances of these algorithms under offline and online knowledge.


\begin{thebibliography}{10}
    
\bibitem{yates09TWC}
J.~Lei, R.~Yates, and L.~Greenstein, ``A generic model for optimizing
  single-hop transmission policy of replenishable sensors,'' {\em IEEE Trans.
  Wireless Commun.}, vol.~8, pp.~547--551, February 2009.

\bibitem{sharma10TWC}
V.~Sharma, U.~Mukherji, V.~Joseph, and S.~Gupta, ``Optimal energy management
  policies for energy harvesting sensor nodes,'' {\em IEEE Trans. Wireless
  Commun.}, vol.~9, pp.~1326--1336, April 2010.

\bibitem{tassiulas10TWC}
M.~Gatzianas, L.~Georgiadis, and L.~Tassiulas, ``Control of wireless networks
  with rechargeable batteries,'' {\em IEEE Trans. Wireless Commun.}, vol.~9,
  pp.~581--593, February 2010.

\bibitem{Zhang10ISIT}
C.~Ho and R.~Zhang, ``Optimal energy allocation for wireless communications
  powered by energy harvesters,'' in {\em IEEE ISIT}, June 2010.

\bibitem{jing10ciss}
J.~Yang and S.~Ulukus, ``Transmission completion time minimization in an energy
  harvesting system,'' in {\em CISS}, March 2010.

\bibitem{tcom-submit}
J.~Yang and S.~Ulukus, ``Optimal packet scheduling in an energy harvesting
  communication system,'' {\em IEEE Trans. Comm.}, submitted June 2010.
\newblock Also available at [arXiv:1010.1295].

\bibitem{kaya_subm}
K.~Tutuncuoglu and A.~Yener, ``Optimum transmission policies for battery
  limited energy harvesting nodes,'' {\em IEEE Trans. Wireless Comm.},
  submitted, September 2010.
\newblock Also available at [ar{X}iv:1010.6280].

\bibitem{eytan09}
M.~Zafer and E.~Modiano, ``A calculus approach to energy-efficient data
  transmission with quality-of-service constraints,'' {\em IEEE/ACM Trans. on
  Networking}, vol.~17, pp.~898--911, June 2009.

\bibitem{fu03TON}
A.~Fu, E.~Modiano, and J.~Tsitsiklis, ``Optimal energy allocation and admission
  control for communications sattelites,'' {\em IEEE/ACM Trans. on Networking},
  vol.~11, pp.~488--500, June 2003.

\bibitem{alvinfu06TWC}
A.~Fu, E.~Modiano, and J.~N. Tsitsiklis, ``Optimal transmission scheduling over
  a fading channel with energy and deadline constraints,'' {\em IEEE Trans.
  Wireless Commun.}, vol.~5, pp.~630--641, March 2006.

\bibitem{elif01INFOCOM}
B.~Prabhakar, E.~Uysal-Biyikoglu, and A.~{El Gamal}, ``Energy-efficient
  transmission over a wireless link via lazy scheduling,'' in {\em IEEE
  INFOCOM}, April 2001.

\bibitem{it_2004}
E.~Uysal-Biyikoglu and A.~{El Gamal}, ``On adaptive transmission for energy
  efficiency in wireless data networks,'' {\em IEEE Trans. Inform. Theory},
  vol.~50, pp.~3081--3094, December 2004.

\bibitem{zafer05alerton}
M.~Zafer and E.~Modiano, ``Continuous time optimal rate control for delay
  constrained data transmission,'' in {\em Allerton Conference}, September
  2005.

\bibitem{BersekasDynProg}
D.~P. Bertsekas, {\em Dynamic Programming and Optimal Control}.
\newblock Athena Scientific, 2007.

\bibitem{BersekasConvexOpt}
D.~P. Bertsekas, {\em Convex Analysis and Optimization}.
\newblock Athena Scientific, 2003.

\bibitem{Goldsmith97}
A.~Goldsmith and P.~Varaiya, ``Capacity of fading channels with channel side
  information,'' {\em IEEE Trans. Inform. Theory}, vol.~43, pp.~1986--1992,
  Nov. 1997.
  
\end{thebibliography}
\end{document}